\documentclass[fleqn,twoside,twocolumn,nofootinbib,showkeys]{Revtex4} 
\usepackage{amsmath}
\usepackage{amsfonts}
\usepackage{amssymb}
\RequirePackage{cmap} 
\RequirePackage[cp1251]{inputenc}
\RequirePackage[T2A]{fontenc}
\RequirePackage{amsmath}
\RequirePackage{amstext}
\RequirePackage{amssymb}
\RequirePackage{bm}
\RequirePackage{graphicx}
\RequirePackage{epstopdf}
\begin{document}
\title{Thermal corrections in Bardeen type regular black holes}
\author{Yawar H.Khan}
\affiliation{Department of Physics National Institute of Technology, Hazratbal, Srinagar, 190006.}
\email{iyhkphy@gmail.com,princeganai@nitsri.com}
\author{Prince A.Ganai}
\affiliation{Department of Physics National Institute of Technology, Hazratbal, Srinagar, 190006.}

\email{iyhkphy@gmail.com,princeganai@nitsri.com}

\setcounter{page}{1}%

\begin{abstract}
 We carry quantum (thermal) corrections to the thermodynamics of regular black holes in detail. Firstly, we discuss the non-extended phase space thermodynamics of regular black holes. We obtain expressions for various thermodynamic quantities like temperature, entropy, internal energy, Helmholtz free energy, pressure, enthalpy, and Gibbs free energy. Then we introduce quantum corrections to the thermodynamics of Bardeen type (toy model) of regular black holes. The qualitative and quantitative analysis of these corrections is carried by plotting all the corrected thermodynamic quantities. The comparative analysis is then done where the effects of these corrections can be clearly seen.
\end{abstract}


\maketitle

\section{Introduction}
 The existence of singularities seems to be unavoidable for almost all of the physically acceptable solutions of Einstein equations  and this is quite true for almost all of the known exact black hole solutions. These singularities occur due to global hyperbolicity in any kind of matter collapsing in the general relativity regime \cite{1}. The existence of such singularities is beautifully interpreted by Penrose cosmic censorship, which states that these singularities must be furnished by event horizons. There exists no casual connection of interiors of a black hole with the exterior fields as the physics outside is well-regulated \cite{e}. However, there are some attempts to remove these singularities like by Sakharov \cite{q} and Gliner \cite{w}. Later on Bardeen replaced the singularity inside black hole  with de-sitter patch and proposed a new kind of black hole called as `regular black hole (RBH)" \cite{2}. Later on, many more RBH models were proposed \cite{3,4,5,6}. All of these models were later termed as ``Bardeen black holes" or "toy model regular black holes" \cite{7,8}. Some characteristic features of RBH  differentiate them from usual singular solutions. For example, spherically symmetric RBH of Bardeen violates strong energy conditions which results in the regularity of these solutions. In a general context, one can say that RBH  violates some conditions of Hawking-Penrose theorems of singularity \cite{12} which results in the regularity of such solutions. It has been found that  RBHs have some special features. For instance,  one can have RBHs with spherical symmetry like Dymnikova \cite{z,x,c}, Bronnikov \cite{v}, and Hayward \cite{b}. Also, there are RBHs with axial symmetry \cite{18,19}. Some RBHs with rotation can also be found in the literature which in general violate weak energy condition \cite{20,21,22,23}. 

Thermodynamics is an important aspect of black hole physics which started with the breakthrough of Hawking who proved that semi-classically a black hole can radiate \cite{24}, referred to as Hawking Radiation. Hawking radiation for black holes is a result of tunneling at horizons \cite{25,26}. This connection between gravity and thermodynamics (as black hole being a completely gravitational object and Hawking radiations carry thermodynamic quantities with them) was then beautifully covered by Bardeen, who formulated four laws of black hole thermodynamics \cite{27}. This connection was sustained by making proportionality between surface gravity and temperature as well as the surface area of the event horizon and entropy. In thermodynamics of black holes, the temperature is obtained from the first law of black hole thermodynamics. The entropy for the black hole is given by Bekenstein's area law \cite{28,vv}. Since then the thermodynamics has become not only an important aspect in black hole physics but it has proven to be a vital tool in characterizing the physicality and viability of many black hole solutions.

Corrections to the thermodynamics of black holes is an important feature that incorporates or brings stability/instability, criticality, and many more other features. In the black hole, thermodynamics corrections due to fluctuations have now got a prominent place and have become by now a frequent subject of interest.  For instance, the corrections to black hole thermodynamics with matter fields in the background have been studied in \cite{jab,sai}.
The effect of thermal fluctuations on a charged anti-de Sitter (AdS) black hole is an excellent accord of corrections in AdS black holes \cite{tere}. The detailed analysis of the thermodynamics of black holes suggested that the quantum approach at small scales to thermodynamics of black holes is inevitable. This resulted in the corrections to various thermodynamic quantities by quantum effects. One such approach that is worth mention is GUP-corrected thermodynamics for all black objects \cite{naina}. The 
correction (of the form $\alpha\ln A$) to G\"odel black hole has been discussed in \cite{pro}. 
These corrections are necessary for logarithmic in nature  \cite{ka}. 
The effect of quantum corrections to thermodynamics of black holes using Cardy formalism has also been studied in \cite{carlip}. 
The detailed effect of quantum corrections on thermodynamics of black holes is discussed by  Nozari et al \cite{lage}. S. Upadhyay in \cite{rai} has discussed quantum corrections to thermodynamics of quasi-
topological black holes. Such kinds of studies for the Schwarzschild-Beltrami-de Sitter black hole \cite{aaj} and the massive black hole in AdS space \cite{din}  have also been made. The detailed study in this regard of quantum corrections to black holes can be found in  \cite{tab}. The quantum corrections in the black hole thermodynamics started with the pioneering work of Frolov et al. \cite{one}. Some other kinds of corrections to the thermodynamics of RBHs do also exist in literature like simple RBH 
with logarithmic entropy correction  \cite{two}. Some other related work could be found in \cite{apple, mango}.

The thermodynamics for regular class of black holes  have been done in many attempts
\cite{29,30}. But these investigations are concerned mainly with calculating temperature, 
entropy, and specific heat only. In this paper, apart from calculating temperature, entropy 
, and specific heat, we calculate various other thermodynamic quantities like free energy, 
internal energy, enthalpy, and Gibbs free energy. These quantities are important tools for determining the stability and time evolution of black holes. Here, we first give a brief idea of the general class of RBHs. We then  discuss the thermodynamics for this general class of 
RBHs. We further incorporate the corrections to the thermodynamic quantities due to quantum. Some earlier attempts of calculating corrections to the thermodynamics of 
RBHs have been made \cite{kk}. Here, we calculate the quantum corrections to the thermodynamics of a specific class of RBHs also (namely, Hayward black holes and Bardeen black holes).

The plan of the paper is as follows. In section II, we recapitulate the basics of RBHs.  The thermodynamics of  RBH is discussed in section III. Further, in section IV, we study the effects of quantum corrections on the thermodynamics of general black holes.  
We emphasize the specific case of Hayward and  Bardeen black holes in sections V and VI, respectively. We conclude our work in the last section.
\section{Brief Review of RBHs}
Let us start by writing the line element for the typical spherically symmetric RBH as follows,
\begin{equation}
ds^2 = - f(r) dt^2 +\frac{dr^2}{f(r)} + r^2 d\theta^2 +r^2\sin^2\theta d\phi^2,
\end{equation}
where metric function $f(r)$ has the following form:
\begin{equation}
f(r)= 1- \frac{2m(r)}{r}.
\end{equation}
The mass term $m(r)$ given in Ref. \cite{23} is of the form
\begin{equation}
m(r) = \frac{M_0}{(1+(\frac{a}{r})^q)^\frac{p}{q}},\label{m}
\end{equation} 
which guarantees   the asymptotic flatness of spacetime for positive values of
$p$ and $q$. Here,  $M_0$ and $a$ correspond to mass and length parameter, respectively. In fact, $M_0$ stands for Arnowitt-Deser-Misner mass of the Schwarzschild
black hole. The basic characteristics of Bardeen type RBHs is carried in the mass term given in Eq.(\ref{m}). For example:\\
1. de Sitter core is generated for small values of $r$ and for $p=3$ \cite{23,se}.\\
2. For large values of $r$, typically $r\ge r_+$ the
mass function is almost constant $m(r)\approx M_0$, then
we have approximately the Schwarzschild metric.\\
3. To identify Bardeen and Hayward RBHs, one specifies $p=3,q=2$ and $ p=q=3$ in mass term respectively.
\section{Thermodynamics}
In this section, we discuss the thermodynamics of the toy model class of RBHs. 
We calculate various thermodynamic quantities like entropy, free energy, internal energy, 
pressure, enthalpy, and Gibbs energy. The thermodynamics for a particular type of RBHs can be easily found in the literature.  For example, thermodynamics of Bardeen RBH is discussed by    Akbar et al. \cite{qqq}, Myung et al. \cite{we, re, hhh}, and  Dymnikova et al.
\cite{ggg}. Their investigations mainly deal with entropy, temperature, and specific heat. 
However,  we are interested in 
derive almost all thermodynamic quantities of physical interest which are so important in determining the stability and viability of a particular black hole. The thermodynamics of any class of RBHs is more attractive than any other class of black holes as RBHs have a temperature less than Schwarzchild black hole. In contrast to Schwarzchild black hole, RBHs show stability for positive specific heat.

Here, we first determine the Hawking temperature ($T_H$) for general class of RBHs. There are various ways to derive $T_H$. A  beautiful discussion on evaluation of temperature for general class of RBHs is given in Ref. \cite{se}. We evaluate $T_H$ by 
\begin{equation}
T_H =\frac{f'{(r_+)}}{4\pi}=\frac{1-2  (\frac{a}{r_+} )^q}{(4 \pi  r)  ( (\frac{a}{r_+} )^q+1 )}.\label{tem}
\end{equation} 
Here, $r_+$ denotes the greater of the roots of the equation $f(r)=0$. In the limit of $a
 arrow 0$, where we get Schwarzchild black hole, one can easily find that Hawking temperature of RBHs is less than Hawking temperature of singular black holes. This indicate that RBHs are colder than any other type of singular black holes. 

We then estimate the entropy of general class of RBH. The entropy of general class of RBHs follows from Bekesteins area law \cite{vv}. In order to determine the entropy we follow
semi-classical formulation which reads as 
\begin{equation}
S_0= \int \frac{1}{T_H}dm.\label{s}
\end{equation} 
Plugging the values of  $T_H$  (\ref{tem}) and  $m$ (\ref{m}) in (\ref{s}) results
\begin{equation}
S_0=\pi  r_+^2. 
\end{equation} 
Next, we calculate the internal energy for the RBH from the formula  $E = \int T_H dS_0$.
Plugging the values of temperature and entropy in this formula leads to 
\begin{equation}
E =    \frac{3 \pi  r_+^2 \,{}_2F_1 (1,-\frac{1}{q};\frac{q-1}{q};- (\frac{a}{r_+} )^q )-2 \pi  r_+^2}{2 \pi  r_+},
\end{equation}\\
where ${}_2F_1$ refers to hypergeometric functions of second kind and should be implied so throughout the text.
We now compute an important thermodynamical quantity called Helmholtz free energy or free energy which is a measure of useful work obtainable from a closed thermodynamic system. 
The free energy $F$ is defined  as
$F=-\int S_0 dT_H$. In order to estimate $F$  for general class of Bardeen type RBHs, we first calculate 
\begin{eqnarray}
dT_H = \frac{(3 q+1)  (\frac{a}{r_+} )^q+2  (\frac{a}{r_+} )^{2 q}-1}{4 \pi  r_+^2  ( (\frac{a}{r_+} )^q+1 )^2}.
\end{eqnarray}. With this value of $dT_H$ and entropy $S_0$, the Helmholtz free energy turns out to be
\begin{equation}
F =    \frac{-6 \pi  r_+^2 \, {}_2F_1 (1,-\frac{1}{q};\frac{q-1}{q};- (\frac{a}{r_+} )^q )+\frac{3 \pi  r_+^2}{ (\frac{a}{r_+} )^q+1}+2 \pi  r_+^2}{4 \pi  r_+}.
\end{equation}
Here we note that to determine the value of $F$ specifically one has to fix the value of $q$. 

Another important thermodynamic quantity on which we focus is pressure. The pressure in Black hole physics has an important significance as this quantity is directly related to the cosmological constant $\Lambda$. The variation in pressure can give an idea about the 
spacetime curvature at horizon. The pressure in general is given by the relation 
$$P=-\frac{dF}{dV}$$ where $F$ is Helmholtz free energy and $V$ is corresponding volume, which in our case is  $V$ = $\frac{4}{3}\pi r_+^3$.
This relation leads to the following value of pressure for spherically symmetric regular black holes:
\begin{equation}
P = \frac{3 q  (\frac{a}{r_+} )^q+ (\frac{a}{r_+} )^q+2  (\frac{a}{r_+} )^{2 q}-1}{4 \pi  r_+^2  ( (\frac{a}{r_+} )^q+1 )^2}.
\end{equation}
The enthalpy   is an important state function which gives an idea about the energy changes of system. The enthalpy of system also provides an idea about the equilibrium conditions of the system. The enthalpy in black hole thermodynamics gained prominence due to Kastor etal. \cite{kastor} when
thermodynamic variables were included in first-law of black hole thermodynamics along with the cosmological constant. There it is argued  that the mass $M$ of an AdS black hole plays the role of enthalpy of classical thermodynamics. We calculate enthalpy for Bardeen class of RBHs by formula $H=E+PV$. This gives,
\begin{eqnarray}
H &=&  \frac{6  (3 \pi  r_+^2{}_2 F_1 (1,-\frac{1}{q},\frac{q-1}{q};-\frac{a}{r_+} )-2 \pi  r_+^2 )}{12 \pi  r_+}\nonumber \\
&+&\frac{\pi  r_+^2  ((3 q+1)  (\frac{a}{r_+} ){}^q+2  (\frac{a}{r_+} ){}^{2 q}-1 )}{ (12 \pi  r_+ )  ( (\frac{a}{r_+} ){}^q+1 ){}^2}.
\end{eqnarray}
For black holes to have a static boundary to be held at a fixed temperature we need fixed pressure and temperature. In such a scenario, the
relevant thermodynamic potential to be used is naturally the Gibbs potential  or Gibbs free energy.
Once  Helmholtz Free energy, pressure and volume are known it is matter of 
calculation to derive Gibbs free energy  by  the formula $G=F+PV$ as following:
\begin{eqnarray}
G&=&\frac{-6 \pi  r_+^2{}_2 F_1 (1,-\frac{1}{q},\frac{q-1}{q};-\frac{a}{r_+} )+\frac{3 \pi  r_+^2}{ (\frac{a}{r_+} ){}^q+1}+2 \pi  r_+^2}{4 \pi  r_+}\nonumber \\
&+&\frac{(3 q+1)  (\frac{a}{r_+} ){}^q+2  (\frac{a}{r_+} ){}^{2 q}-1}{12  ( (\frac{a}{r_+} ){}^q+1 ){}^2}
\end{eqnarray} 
This completes the relevant equations of states for thermodynamics of toy model class of RBHs. These thermodynamic quantities can be used for analysis of specific black holes by fixing  the value of $q$. Thus for different values of $q$ (i.e. for even and odd $q$), one can have different nature of thermodynamic quantities.
We shall discuss these things in detail in later sections by setting the values of $q$ for Hayward and Bardeen classes of RBHs.
\section{Quantum corrections}
In this section, we shall analyze the effect of quantum corrections to the thermodynamics of the toy model class of RBHs.  Some of the earlier attempts concerning the application of The quantum correction to black hole thermodynamics could also be found in the literature. Here we employ the quantum corrections to the thermodynamics of the toy model class of regular black holes. We start our discussion by introducing the quantum correction parameter $\alpha$ in the entropy equation. The basic essence of this type of corrections follows from the quantum regime. In quantum regime, near the Planck scale, the quantum corrections to gravity are expected to change the manifold structure of spacetime \cite{bak, rama}. This can change the holographic principle as a result of which an evident change in entropy-area law is expected \cite{pour}. It was then
argued first in \cite{tab} and then in  \cite{pour} that corrected entropy-area relation could be written as 
$$ S = S_0 + \alpha lnA + \gamma_1 A^{-1} + \gamma_2A^{-2} +.........$$
where $\alpha, \gamma_1, \gamma_2 $ depend on the type of model under consideration. Therefore one gets (up to
first order)
\begin{eqnarray}
S = S_0 + \alpha \log [A] =\pi  r_+^2+\alpha \log  (4 \pi  r_+^2 ). \label{ent}
\end{eqnarray} 
Due to the correction term in entropy, every thermodynamical quantity will get the modification.
We first start computing internal energy $E$. 
The corrected value of internal energy  can be obtained by 
formula  $E_C=\int T_H dS$  as
\begin{eqnarray}
E_C &= &\frac{3 \pi  r_+^2 \, {}_2F_1 (1,-\frac{1}{q};\frac{q-1}{q};- (\frac{a}{r_+} )^q )-2 \pi  r_+^2 }{2 \pi  r_+}\nonumber \\
&-&\frac{3 \alpha  \, {}_2F_1 (1,\frac{1}{q};1+\frac{1}{q};- (\frac{a}{r_+} )^q )+2 \alpha }{2 \pi  r_+}.\label{ec}
\end{eqnarray} 
This  gives the expression for the quantum corrected internal energy for toy model class of RBHs. The detailed analysis of these corrections of internal energy would be calculated for different classes of RBHs in later sections.

Furthermore, we derive the the quantum corrected Helmholtz free energy denoted by 
relation  $F_C = \int S dT_H$. Here,  we get
\begin{eqnarray}
F_C&=& \frac{-6 \pi  r_+^2 \, {}_2F_1 (1,-\frac{1}{q};\frac{q-1}{q};- (\frac{a}{r_+} )^q )+\frac{3 \pi  r_+^2}{ (\frac{a}{r_+} )^q+1}+2 \pi  r_+^2}{4 \pi  r_+} \nonumber\\
&+& \frac{6 \alpha  \, {}_2F_1 (1,\frac{1}{q};1+\frac{1}{q};- (\frac{a}{r_+} )^q )+\frac{3 \alpha  \log  (4 \pi  r_+^2 ) }{ (\frac{a}{r_+} )^q+1}}{4 \pi r_+}\nonumber \\ 
&-&
\frac{4 \alpha -2 \alpha  \log  (4 \pi  r_+^2 )}{4 \pi r_+}.\label{fc}
\end{eqnarray}
This equation represents the quantum corrected Helmholtz Free energy for toy model class of RBHs. These corrections would be evident once we interpret this $F_C$ for different classes of black holes simply by setting different values of $q$.

The the quantum correction can be incorporated in the pressure equation 
by equation $P=-\frac{dF_C}{dV}$ as,
\begin{equation}
P_C=\frac{ (3 q  (\frac{a}{r_+} )^q+ (\frac{a}{r_+} )^q+2  (\frac{a}{r_+} )^{2 q}-1 )  (\alpha  \log  (4 \pi  r_+^2 )+\pi  r_+^2 )}{4 \pi ^2 r_+^4  ( (\frac{a}{r_+} )^q+1 )^2}.\label{p}
\end{equation}
The above equation gives the modified value of pressure and could be recasts for various classes of RBHs simply by substituting different values of $q$. 

The expression for corrected value of enthalpy would follow from relation $H_C = E_C+P_CV $. Exploiting the value of $E_C$, $P_C$ and volume $V$ = $\frac{4}{3}\pi r_+^3$, we get the following corrected enthalpy:
\begin{eqnarray}
H_C &=& \frac{-2 \pi r_+^2+2\alpha+3\pi r_+^2 {}_2F_1[1,-\frac{1}{q},\frac{-1+q}{q},-(\frac{a}{r_+})^q]}{2\pi r_+} \nonumber\\
&-&\frac{\alpha {}_2F_1[1,\frac{1}{q},\frac{1+q}{q},-(\frac{a}{r_+})^q]}{4\pi r_+}\nonumber \\
&+&\frac{ -1+(1+3q)(\frac{a}{r_+})^q+2(\frac{a}{r_+})^2q}{12 \pi r_+^2(1+(\frac{a}{r_+})^q)^2} \nonumber \\
&+&\frac{ \pi r_+^2+\alpha \log[4 \pi r_+^2] }{12 \pi r_+^2(1+(\frac{a}{r_+})^q)^2}.\label{hc}
\end{eqnarray} 
Next, we calculate quantum modified Gibbs free energy for the Bardeen class of RBHs. To do so, we  use relation $G_C=F_C+P_CV$  and obtain,
\begin{eqnarray}
G_C &=& \frac{-6 \pi  r_+^2 \, {}_2F_1 (1,-\frac{1}{q};\frac{q-1}{q};- (\frac{a}{r_+} )^q )+2 \pi  r_+^2}{4 \pi  r_+}\nonumber\\
&+& \frac{3  (\alpha  \log  (4 \pi  r_+^2 )+\pi  r_+^2 )}{4 \pi  r_+ ( (\frac{a}{r_+} )^q+1 )}\nonumber \\
&+& \frac{-4 \alpha -2 \alpha  \log  (4 \pi  r_+^2 )+6 \alpha  \, {}_2F_1 (1,\frac{1}{q};1+\frac{1}{q};- (\frac{a}{r_+} )^q )}{4 \pi r_+}\nonumber\\
&+&\frac{ (3 q  (\frac{a}{r_+} )^q+(\frac{a}{r_+} )^q+2(\frac{a}{r_+} )^{2 q}-1)(\alpha  \log  (4 \pi  r_+^2 )+\pi  r_+^2)}{12 \pi r_+((\frac{a}{r_+})^q+1)^2}.\label{gc}
\end{eqnarray}
This section completes the calculation of the quantum corrected thermodynamics for toy model class of RBHs. The effect of these corrections would be determined once we go for specific  class of RBHs. All these effects would be controlled by nature of the integral values of $\alpha$ which can take any integral value. We expect that the effect of the corrections would be logarithmic in nature, by Kaul and Majmudar \cite{ka}. These corrections can induce various thermodynamic features into any class of black holes including RBHs like stability, instability, criticality and altering heat capacity etc.
\section{Hayward Black Holes}
\begin{figure}[htb]
	\begin{center}$
		\begin{array}{c }
		\includegraphics[width=80 mm]{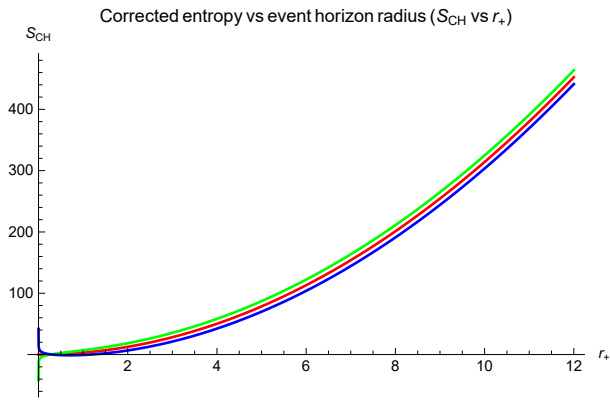}  
		\end{array}$
	\end{center}
	\caption{Entropy versus $r_+$. Red, green and blue curves corresponds to $\alpha =0, \alpha
		=1$ and  $\alpha=-1.5$, respectively.}
	\label{fig1}
\end{figure}

In the previous section, we calculated quantum corrections to the thermodynamics of toy model class of RBHs. To study the effects of  these correction,  we plot resulting modified thermodynamic quantities against various values of correction parameter $\alpha$. First,   we set $q=3$ which corresponds to the Hayward black holes. 
The effect of quantum corrections would be maximum at small values of $r_+$. In fact, at large distances, these corrections have almost null consequences.
By simply putting $q=3$ in (\ref{tem}), the temperature for Hayward RBHs is given by 
\begin{equation}
T=\frac{1-\frac{2 a^3}{r_+^3}}{4 \pi  r_+  (\frac{a^3}{r_+^3}+1)}.
\end{equation} 
We now plot entropy (\ref{ent}) against $r_+$ for various values of $\alpha$ in Fig. (\ref{fig1}).
From this plot, we notice that the effect of quantum corrections is notable for small values of $r_+$ which is quite obvious. For positive values of $\alpha$  entropy decreases which infers instability; however for negative $\alpha$ we can see that entropy tends to increase asymptotically which is clear sign of stability. This justifies  that quantum corrections corresponding to  negative value of correction parameter induce stability in Hayward type of RBHs.

\begin{figure}[htb]
	\begin{center}$
		\begin{array}{c }
		\includegraphics[width=80 mm]{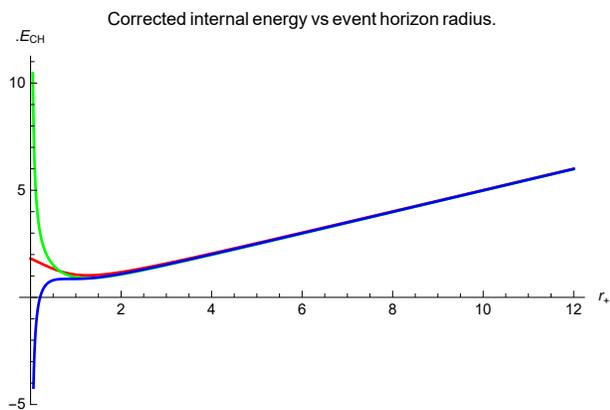}  
		\end{array}$
	\end{center}
	\caption{Internal energy versus $r_+$. Red, green and blue curves corresponds to $\alpha =0, \alpha
		=1$ and  $\alpha=-1.5$, respectively.}
	\label{fig2}
\end{figure}

The expression for internal energy for Hayward black holes, $E_{CH}$, can be obtained from 
Eq. (\ref{ec}) 
by plugging $q=3$  as follows,
\begin{eqnarray}
E_{CH} &=&\frac{ 3 \pi  r_+^2 \, {}_2F_1 (-\frac{1}{3},1;\frac{2}{3};-\frac{a^3}{r_+^3} )-2 \pi  r_+^2}{2 \pi  r_+}\nonumber\\
&-&\frac{3 \alpha  \, {}_2F_1 (\frac{1}{3},1;\frac{4}{3};-\frac{a^3}{r_+^3} )+2 \alpha }{2 \pi  r_+}.
\end{eqnarray} 
In order to do comparative analysis, we plot a graph \ref{fig2}   for $E_{CH}$ vs $\alpha$.
From the plot, it is obvious that for negative value of $\alpha$ internal energy for small black holes decreases. In contrast, for positive values of $\alpha$ it takes asymptotically   positive value. This is in agreement of first law of black hole thermodynamics as we have already seen that for negative value of $\alpha$ entropy increases. Thus, even after incorporating the   the quantum correction,  the first laws of black hole thermodynamics holds. The decrease in internal energy of Hayward black holes due quantum corrections leads to stability of the black holes. 

In order to estimate the effect of corrections on the free energy of Hayward type of black holes, we first set $q=3$ in Eq. (\ref{fc}). This leads to,
\begin{eqnarray}
F_{CH}  &=&\frac{-6 \pi  r_+^2 \, {}_2F_1 (-\frac{1}{3},1;\frac{2}{3};-\frac{a^3}{r_+^3} )+2 \pi  r_+^2}{4 \pi r_+}\nonumber\\
&+&\frac{6 \alpha  \, {}_2F_1 (\frac{1}{3},1;\frac{4}{3};-\frac{a^3}{r_+^3} )}{4 \pi r_+}
+ \frac{3 r_+^3  (\alpha  \log  (4 \pi  r_+^2 )+\pi  r_+^2 )}{4 \pi r_+ (a^3+r_+^3)}  \nonumber\\
&-&\frac{4 \alpha -2 \alpha  \log  (4 \pi  r_+^2 )}{4 \pi  r_+}.
\end{eqnarray}
Here we note that after setting $\alpha =0$ we obtain the original expression for free energy for Hayward type of black holes. 
\begin{figure}[htb]
	\begin{center}$
		\begin{array}{c }
		\includegraphics[width=80 mm]{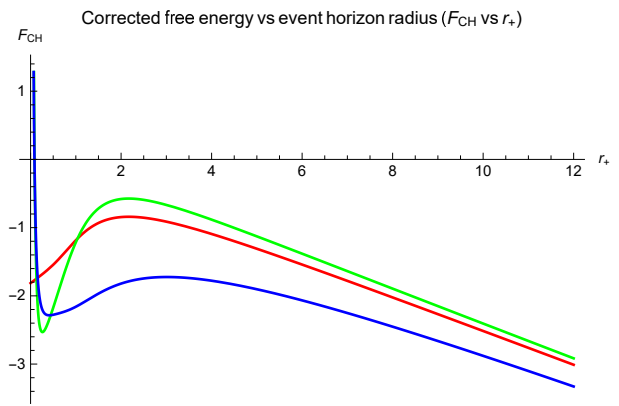}  
		\end{array}$
	\end{center}
	\caption{Helmholtz free energy versus $r_+$. Red, green and blue curves corresponds to $\alpha =0, \alpha
		=1$ and  $\alpha=-1.5$, respectively.}
	\label{fig3}
\end{figure} 
In order to study the effect of the quantum on the free energy of Hayward black hole,
we plot Fig.  \ref{fig3}. From the figure, one can clearly see that the positive values of $\alpha$ increases the Helmholtz free energy   which indicates to instability. In contrast, negative $\alpha$  decreases   Helmholtz free energy. For smaller black holes, Helmholtz free energy has positive asymptotic value, which indicates instability. This plot helps us  to identify the region of instability. The existence of such an unusual behavior of Helmholtz free energy curve for $r_+$ $ arrow 0$  is due to the overcome of quantum effects by tidal forces at very small values of $r_+$. This will also get clear from the   pressure. 

\begin{figure}[htb]
	\begin{center}$
		\begin{array}{c }
		\includegraphics[width=80 mm]{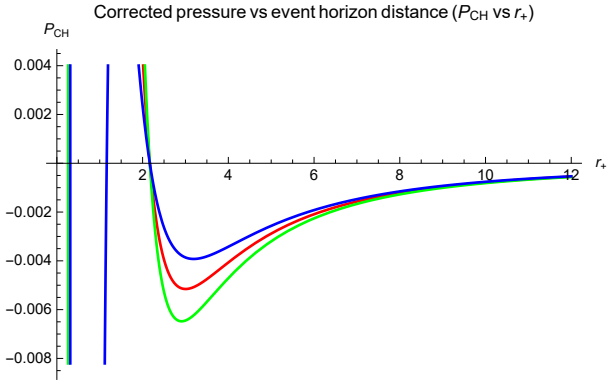}  
		\end{array}$
	\end{center}
	\caption{ Pressure versus $r_+$. Red, green and blue curves corresponds to $\alpha =0, \alpha
		=1$ and  $\alpha=-1.5$, respectively.}
	\label{fig4}
\end{figure}

The the quantum corrected  pressure for Hayward type of RBHs, $P_{CH}$, can be obtained  from Eq. (\ref{p}) by substituting $q=3$. Thus, 
\begin{equation}
P_{CH} = -\frac{ (-2 a^6-10 a^3 r_+^3+r_+^6 )  (\alpha  \log  (4 \pi  r_+^2 )+\pi  r_+^2 )}{4 \pi ^2 r_+^4  (a^3+r_+^3 )^2}.
\end{equation}
The choice of value of $\alpha$ will give the detailed effect of corrections on the pressure for Hayward black holes. We plot $P_{CH}$ vs $r_+$ in the   figure \ref{fig4}.
From the plot \ref{fig4}, it is obvious that there exist  two regions. At $r_+ = 1$  pressure is a discontinuous function. The region  $r_+ > 1$  shows a typical pressure curve for black holes. In this region, the effect of The quantum  gravity corrections is   clearly evident. For positive values of $\alpha$ (i.e. $\alpha =1$) the pressure becomes more negative. For negative values of $\alpha$ (i.e. $\alpha=-1.5$)  the pressure   shifts towards positive value. This justifies  that negative $\alpha$ increases stability of the black holes, which follows the case of entropy, internal energy and free energy. In the region of $r_+< 1$,  the pressure curves for different values of $\alpha$ do not change with $r_+$. This shows that in the region of $r_+< 1$ no sort of effects, fluctuations or corrections can overcome the tidal forces which are associated with pressure of black hole. Since  the pressure of  the black hole is related to the cosmological constant of that black hole, we conclude that the cosmological constant determines the type and strength of correction which would get affected at any value of $r_+$.

\begin{figure}[htb]
	\begin{center}$
		\begin{array}{c }
		\includegraphics[width=80 mm]{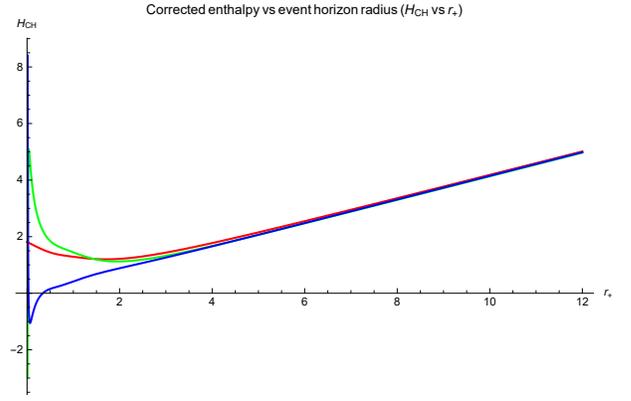}  
		\end{array}$
	\end{center}
	\caption{ Variation  of $H_{CH}$ w.r.t $r_+$ for different values of $\alpha$. Red, green and blue curves corresponds to $\alpha =0, \alpha
		=1$ and  $\alpha=-1.5$, respectively.}
	\label{fig5}
\end{figure}

We now analyze the effect of correction parameter $\alpha$ on enthalpy of Hayward black holes. The expression for corrected enthalpy $H_{CH}$ is given by 
\begin{eqnarray}
H_{CH}&=&-\frac{6 \pi  r_+^2 \, 2F1 (-\frac{1}{3},1;\frac{2}{3};-\frac{a^3}{r_+^3} )-2 \pi  r_+^2}
{4 \pi r_+}\nonumber\\
&+&\frac{6 \alpha  \, 2F1 (\frac{1}{3},1;\frac{4}{3};-\frac{a^3}{r_+^3} )}{4 
	\pi r_+}
+ \frac{3 r_+^3  (\alpha  \log  (4 \pi  r_+^2 )+\pi  r_+^2 )}
{4 \pi r_+({a^3+r_+^3})}\nonumber\\
&-&\frac{4 \alpha -2 \alpha  \log  (4 \pi  r_+^2 )}{4 \pi  r_+}.
\end{eqnarray}  
The plot of corrected enthalpy with respect to horizon radius is given in figure \ref{fig5}. 
This is  clear from plot that negative value of $\alpha$ given by blue line decreases enthalpy and hence induces stability  for $r_+$ tends to zero. Again a kink in the graph of negative values of $\alpha$ is attributed to that overcome of quantum effects by tidal forces. The asymptotic shift of blue line at $r_+\rightarrow 0$  is also a kind of unusual behavior. This can also justified by the overcome of quantum effects by tidal forces. 

\begin{figure}[htb]
	\begin{center}$
		\begin{array}{c }
		\includegraphics[width=80 mm]{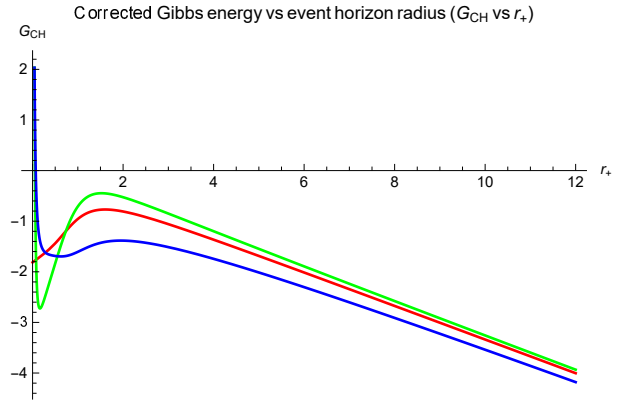}  
		\end{array}$
	\end{center}
	\caption{ Variation  of  $G_{CH}$ w.r.t $r_+$ for different value of $\alpha$. Red, green and blue curves corresponds to $\alpha =0, \alpha
		=1$ and  $\alpha=-1.5$, respectively.}
	\label{fig6}
\end{figure}

Now, we analyze the effect of corrections on the Gibbs free energy of Hayward black holes. The corrected Gibbs free energy $G_{CH}$ is calculated by
\begin{eqnarray}
G_{CH} &=&\frac{-18 \pi  r_+^2 \, {}_2F_1 (-\frac{1}{3},1;\frac{2}{3};-\frac{a^3}{r_+^3} )-8 \pi  r_+^2}{12 \pi r_+}\nonumber\\
	&+&\frac{18 \alpha  \, {}_2F_1 (\frac{1}{3},1;\frac{4}{3};-\frac{a^3}{r_+^3} )}{12 \pi r_+}
+\frac{15 r_+^3  (\alpha  \log  (4 \pi  r_+^2 )+\pi  r_+^2 )}{12 \pi r_+ (a^3+r_+^3 )}\nonumber\\
&-&\frac{9 r_+^6  (\alpha  \log  (4 \pi  r_+^2 )+\pi  r_+^2 )}{12 \pi  r_+ (a^3+r_+^3 )^2} \nonumber\\
&-& \frac{12 \alpha +4 \alpha  \log  (4 \pi  r_+^2 )}{12 \pi  r_+}.
\end{eqnarray}
To study the behavior of the corrected Gibbs free energy, we plot Fig. \ref{fig6}.
From the plot,  we see that  without quantum correction Gibbs free energy is negative for Hayward black holes. Then for positive values of $\alpha$  makes Gibbs energy less negative. However, for negative value of $\alpha$, Gibbs energy becomes more negative value. This is a clear sign of stability. Also, for negative $\alpha$, the Gibbs free energy takes positive asymptotic value for $r_+ \rightarrow 0$. For large black holes, we see that the quantum corrections have almost null effect. The unusual behavior of corrected potential  has been attributed to   overcome the the quantum corrections by tidal forces. The presence of tidal forces has already been seen in the pressure plot \ref{fig4}.
\section{Bardeen Black Holes}
\begin{figure}[htb]
	\begin{center}$
		\begin{array}{c }
		\includegraphics[width=80 mm]{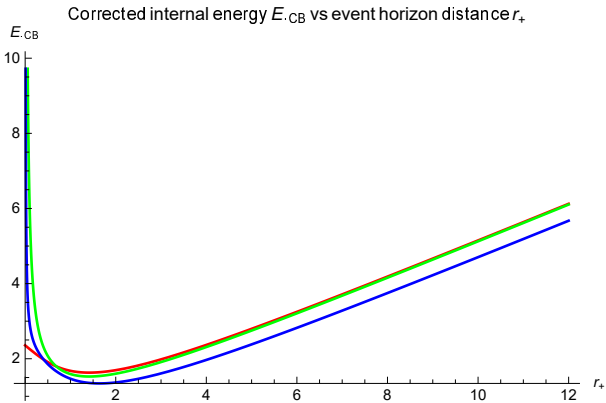}  
		\end{array}$
	\end{center}
	\caption{ Internal energy  w.r.t  $r_+$ for different value of $\alpha$. Red, green and blue curves corresponds to $\alpha =0, \alpha = 1$ and  $\alpha=-1.5$, respectively.}
	\label{fig7}
\end{figure}

The  Bardeen class of RBHs started with the pioneer work of Bardeen \cite{27}.  In order to study  thermodynamics of Bardeen black holes 
we set $q=2$ in the thermodynamics of toy model class of RBHs.
We start by writing the expression of Hawking temperature for Bardeen class of regular black as follows,
\begin{eqnarray}
T_B = \frac{1-\frac{2 a^2}{r_+^2}}{4 \pi  r_+  (\frac{a^2}{r_+^2}+1 )}.
\end{eqnarray}
We then compute  corrected internal energy of Bardeen type RBHs by substituting $q=2$ in the Eq. (\ref{ec}) as,
\begin{equation}
E_{CB} =  \frac{ 3}{2  }    r_+ \, _2F_1 (-\frac{1}{2},1;\frac{1}{2};-\frac{a^2}{r_+^2} )-\frac{3 \alpha  \tan ^{-1} (\frac{a}{r_+} )}{2 \pi a}+\frac{  \alpha }{\pi r_+}-   r_+.  
\end{equation}
The plot \ref{fig7} describes the detailed effect of correction parameter on internal energy 
of Bardeen type RBHs.  
Here, we can see that both positive as well as negative values of $\alpha$ increase internal energy when $r_+ \rightarrow 0$. Contrary to Hayward  RBHs case,  the internal energy of Bardeen type black holes at $r_+ \rightarrow 0$ takes positive asymptotic value for negative  $\alpha$ have a shift in internal energy curve for negative $\alpha$. This difference is simply attributed to the even power of $q$ (i.e. $q=2$ in case of Bardeen class of RBHs). This clearly shows that the quantum corrections induce instability in Bardeen class of RBHs.

\begin{figure}[htb]
	\begin{center}$
		\begin{array}{c }
		\includegraphics[width=80 mm]{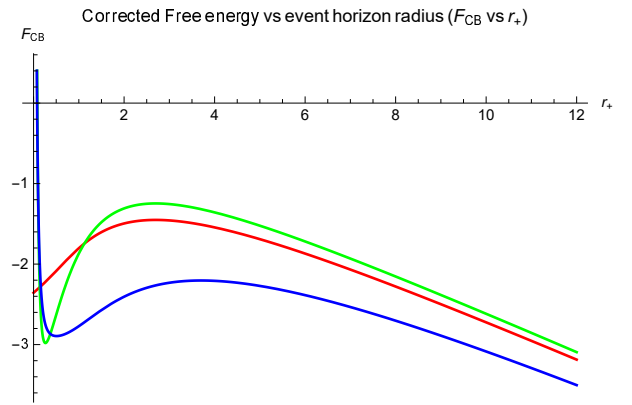}  
		\end{array}$
	\end{center}
	\caption{ Variation  of $F_{CB}$  w.r.t  $r_+$ for different value of $\alpha$. Red, green and blue curves corresponds to $\alpha =0, \alpha
		=1$ and  $\alpha=-1.5$, respectively.}
	\label{fig8}
\end{figure}

Next, we proceed to evaluate the  the quantum effects on Helmholtz free energy of Bardeen   black holes. The corrected Helmholtz free energy, $F_{CB}$ can be estimated by,
\begin{eqnarray}
F_{CB} &=& \frac{-4 \alpha   (a^2+r_+^2 )+\alpha   (r_+^2-2 a^2 ) \log  (4 \pi  r_+^2 )}{4 \pi  r_+(a^2+r_+^2)}\nonumber\\
&-&\frac{\pi  r_+^2  (4 a^2+r_+^2 )}{4 \pi  r_+(a^2+r_+^2)}\nonumber\\
&+&\frac{3}{2 \pi   } (\frac{ \alpha  }{a}-  \pi  a   ) \tan ^{-1} (\frac{a}{r_+} ).
\end{eqnarray} 
The detailed variation of $F_{CB}$ can  be depicted from the figure \ref{fig8}. 
Here, we  find that corresponding to positive value of correction parameter  Helmholtz free energy for large sized black holes becomes less negative than the uncorrected one. 
For smaller black holes (i.e. $r_+<1$),  the behavior is attributed to tidal force dominance. For negative value of correction parameter, we see the free energy of larger black holes (i.e. 
$r_+ > 1$) 
becomes more negative. This indicates the  sign of stability.    

\begin{figure}[htb]
	\begin{center}$
		\begin{array}{c }
		\includegraphics[width=80 mm]{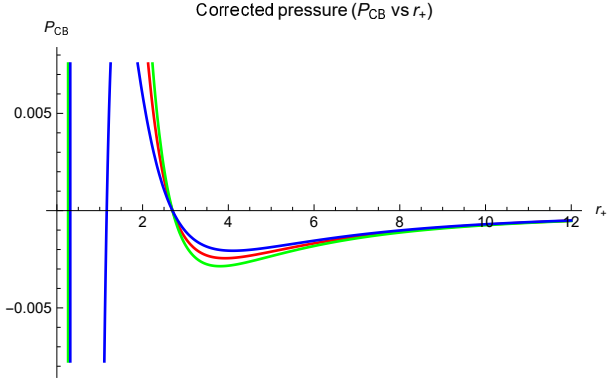}  
		\end{array}$
	\end{center}
	\caption{ Variation  of $P_{CB}$  w.r.t  $r_+$ for different value of $\alpha$. Red, green and blue curves corresponds to $\alpha =0, \alpha
		=1$ and  $\alpha=-1.5$, respectively.}
	\label{fig9}
\end{figure}

\begin{figure}[htb]
	\begin{center}$
		\begin{array}{c }
		\includegraphics[width=80 mm]{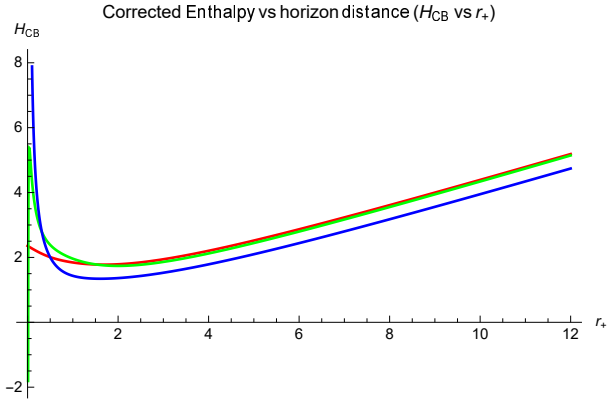}  
		\end{array}$
	\end{center}
	\caption{ Variation  of $H_{CB}$  w.r.t  $r_+$ for different value of $\alpha$. Red, green and blue curves corresponds to $\alpha =0, \alpha
		=1$ and  $\alpha=-1.5$, respectively.}
	\label{fig10}
\end{figure}

The corrected pressure for Bardeen black holes is calculated by setting $q=2$ in Eq. (\ref{p}) 
as follows,
\begin{equation}
P_{CB}=\frac{ (\frac{2 a^4}{r_+^4}+\frac{7 a^2}{r_+^2}-1 )  (\alpha  \log  (4 \pi  r_+^2 )+\pi  r_+^2 )}{4 \pi ^2 r_+^4  (\frac{a^2}{r_+^2}+1 )^2}.
\end{equation}
The  graph for $P_{CB}$ vs $r_+$ is plotted in  Fig. \ref{fig9}. 
The effect of the quantum corrections on pressure could be easily visualized from
the figure. There exist two different regions; one for $r_+<1$ and other for $r_+>1$. In the former region pressure shows unusual behavior as one can see the pressure becomes divergent for different values of $\alpha$.   For  other region, we can see the behavior of pressure is 
quite typical. The  pressure has minimum value near $r_+=4$. For positive value of $\alpha$ (i.e. $\alpha=1 $)  pressure becomes  more negative valued. This is an evidence of instability. For negative value of $\alpha$ (i.e. $\alpha=-1.5$), the pressure becomes relatively more positive valued. 
Next, to calculate the corrected enthalpy $H_{CB}$ for Bardeen type black hole we substitute $q=2$ in Eq.(\ref{hc}). This results,
\begin{eqnarray}
H_{CB} &=& \frac{1}{2 \pi r_+}  (3 \pi  r_+^2 \, _2F_1 (-\frac{1}{2},1;\frac{1}{2};-\frac{a^2}{r_+^2} )\nonumber\\
&-&\frac{3 \alpha  r_+ \tan ^{-1} (\frac{a}{r_+} )}{a}+2 \alpha -2 \pi  r_+^2 )\nonumber\\
&+&\frac{ (2 a^4+7 a^2 r_+^2-r_+^4 )  (\alpha  \log  (4 \pi  r_+^2 )+\pi  r_+^2 )}{12 \pi  r_+ (a^2+r_+^2 )^2}.
\end{eqnarray}
The behavior of enthalpy  calculated above   for different values of correction parameter can be seen in the plot  (\ref{fig10}).
The red curve corresponds to the enthalpy  without the quantum corrections. 
From the figure,  we notice that for small black holes, the  enthalpy of Bardeen black holes  due to  positive value  of $\alpha$
increases which is a clear sign of instability. The negative values of $\alpha$  decreases   the enthalpy. As we know the enthalpy is attributed to the mass of black hole thus decrease in entropy is a clear sign of stability. We thus conclude that negative values of correction parameter induces stability by decreasing enthalpy. 

\begin{figure}[htb]
	\begin{center}$
		\begin{array}{c }
		\includegraphics[width=80 mm]{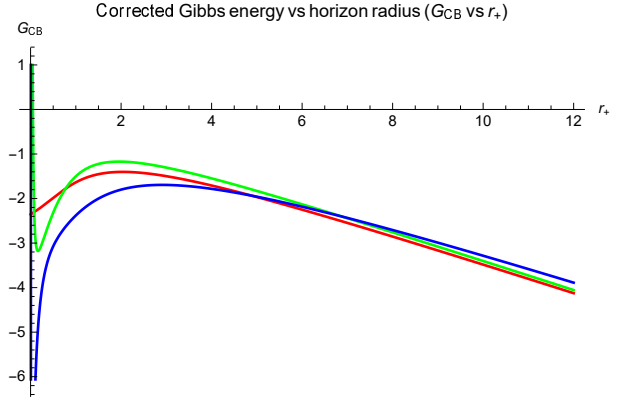}  
		\end{array}$
	\end{center}
	\caption{ Variation  of $G_{CH}$  w.r.t.  $r_+$ for different value of $\alpha$. Red, green and blue curves corresponds to $\alpha =0, \alpha
		=1$ and  $\alpha=-1.5$, respectively.}
	\label{fig11}
\end{figure}

Finally, we  derive expression for the corrected Gibbs energy of Bardeen type of regular black 
holes.  In this regard,  we plug the value $q=2$ in Eq. (\ref{gc}) which yields,
\begin{eqnarray} 
G_{CB}&=& \frac{3 \alpha  \, {}_2F_1 (\frac{1}{3},1;\frac{4}{3};-\frac{a^3}{r_+^3} )}
{2 \pi r_+ }-\frac{ 3   r_+ \, 2F1 (-\frac{1}{3},1;\frac{2}{3};-\frac{a^3}{r_+^3} )}
{2  }\nonumber\\
&+& \frac{5 r_+^2  (\alpha  \log  (4 \pi  r_+^2 )+\pi  r_+^2 )}{4 \pi 
	(a^3+r_+^3)}\nonumber\\
&-&\frac{3 r_+^5  (\alpha  \log  (4 \pi  r_+^2 )+\pi  r_+^2 )}{4 \pi 
	 (a^3+r_+^3 )^2}\nonumber \\
&-&\frac{3 \alpha +  \alpha  \log  (4 \pi  r_+^2 )-2 \pi  r_
	+^2}{3 \pi r_+}.
\end{eqnarray}
We now plot $G_{CB}$ vs $r_+$ and see the effect of corrections on the Gibbs free energy of Bardeen type of RBHs in Figure \ref{fig11}. 
The uncorrected Gibbs free energy for Bardeen type RBHs is negative, which 
confirms the stability of black holes. The the quantum does not affect the stability
of larger black holes. However, in the limit $r_+ \rightarrow 0$ limit, the Gibbs  free energy
takes positive asymptotic values.     This is clear evidence of instability. For the
small black holes, the  negative value of $\alpha$ induce more negativity in Gibbs energy which makes the black holes more stable. This is in agreement for the other cases of black holes mentioned above too.

This completes our analysis of the effect of the quantum corrections on the Bardeen class of RBHs. 
We conclude here that the negative value of $\alpha$ again proved to be a prominent factor in determining
and bringing stability to the Bardeen class of RBHs.

\section{Conclusions}
In this article,  we have carried the thermodynamics of the toy model class of RBHs in detail. The essence of this detailed calculation of thermodynamics was, to sum up, all the relevant calculations and analysis about the thermodynamics of the toy model class of RBHs. We started our discussion by calculating entropy and temperature for the general class of RBHs. Following from area-law,  the entropy for the general class of RBHs is seen to be in par with the generalized second law. We have used the semi-classical formulation to calculate entropy. The Hawking temperature for toy model class of RBHs, on the other hand, can be calculated by various approaches\cite{se}.  We have calculated Hawking temperature followed by calculating internal energy for the toy model class of RBHs. Once entropy and temperature are known, it is a matter of calculation to estimate Helmholtz free energy of the system which gives an idea about surplus energy present in any thermodynamic system. Subsequently, we have calculated pressure, another important thermodynamic quantity. Generally, the pressure of black holes is related to the cosmological constant. This quantity needs special attention as the strength of interaction as well as the presence of different forces is directly related to pressure and in turn cosmological constant. To verify the first law of black hole thermodynamics we need some other thermodynamic state functions too. In this regard, we have calculated enthalpy and Gibbs free energy for the toy model class of RBHs. These thermodynamic potentials, Gibbs Free energy, and Helmholtz Free energy, directly give us an idea about the stability of the black hole thermodynamic system and its evolution in time.  

Furthermore,  we have studied the effect of the quantum on toy model class RBHs. We have considered the corrections up to leading-order only. These corrections,  which are prominent for very small values of $r_+$, were subjected just to bring stability to some thermodynamic quantities. The effect of these corrections on more specific classes of RBHs was evident once we have fixed the value of $q$ for appropriate RBHs.

We have analyzed, in detail, the effect of the quantum corrections on Hayward type RBHs by fixing $q=3$. Here we have found that the correction term with only negative correction parameter increases the entropy of the system. Thus,  we can easily claim that a negative value of $\alpha$ causes stability to tot the system. Then the corrected value of internal energy for Hayward type RBHs was calculated by substituting $q= 3$ in the general case. Here we have seen that the correction due to the quantum is also in agreement with the first law of black hole mechanics as for the negative value of $\alpha$ internal energy decreased. The corrections to the Helmholtz free energy have also been calculated and the effect of correction was expectedly found to decrease the function for the negative value of $\alpha$. The variation of free energy versus horizon radius for different values of 
correction parameter shows deviation from expected behavior at very small values of $r_+$
(i.e $r_+\rightarrow 0$). The corrections to the pressure for the Hayward class of RBHs is also estimated. From the plot, we found that the effect of these corrections was evident only for a larger horizon radius ($r_+> 1$). For region below $r_+<$ 1, these corrections do not play a significant role and the pressure remains constant. Moreover, we have calculated the corrections to the enthalpy of Hayward black holes and found the correction terms expectedly decrease the enthalpy for the negative value of correction parameter $\alpha$. Finally, we have evaluated corrections to Gibbs free energy, where we have found that the quantum corrections with a negative value of $\alpha$ decrease the Gibbs free energy and increases for positive $\alpha$.
We have discussed the quantum corrections to the Bardeen type of RBHs. Here we observed that for the negative value of correction parameter there is an increase in entropy. However, in contrast to the virtue of the first law of black hole mechanics, the internal energy decrease for a negative value of $\alpha$. This is due to the even power $q$. The decrease in Helmholtz free energy for the Bardeen class of RBHs for a negative value of $\alpha$  indicates stability for the system. We have derived the quantum corrections on the pressure of the Bardeen class of RBHs also. Here, we have noticed that like Hayward class of black holes there exist two regions in the corrected pressure curve for Bardeen black holes also. In the region for small horizon radius (i.e. $r_+ < 1$), the pressure remains unaffected by kind of corrections. In the region for a larger horizon radius (i.e.  $r_+> 1$), the pressure shows a typical behavior black hole system. There is an increase in pressure for the negative value of the correction parameter $\alpha$. The behavior of pressure towards the quantum corrections for Bardeen type RBHs is more or less similar to that of Hayward type black holes. This is again due to the dependence of various thermodynamic quantities on integer $q$ which is even for the Bardeen class of RBHs. Also, the behavior of corrected enthalpy and Gibbs free energy  
have found the same to that for Hayward class of RBHs.


\end{document}